\documentclass[aps,prd,preprintnumbers,nofootinbib]{revtex4}
\usepackage[dvips]{graphicx}
\usepackage{amsmath}
\usepackage{bm,latexsym,amsmath,amsthm,amssymb,amsfonts,color}

\usepackage[normalem]{ulem}

\begin{document}

\title{S-matrix Unitarity and Renormalizability in Higher Derivative Theories}
\preprint{KOBE-COSMO-18-03}


\author{Yugo Abe$^{1}$, Takeo Inami$^{2,3}$, Keisuke Izumi$^{4,5}$, Tomotaka Kitamura$^{6}$ and Toshifumi Noumi$^{7}$}
\affiliation{${}^1$National Institute of Technology, Miyakonojo College, Miyakonojo 885-8567,Japan}
\affiliation{${}^2$Theoretical Research Division, Nishina Center, 
RIKEN, Wako 351-0198, Japan}
\affiliation{${}^3$Department of Physics, Sungkyunkwan University, 
Suwon 16419, Republic of Korea}
\affiliation{${}^4$Kobayashi-Maskawa Institute, Nagoya University, Nagoya 464-8602, Japan}
\affiliation{${}^5$Department of Mathematics, Nagoya University, Nagoya 464-8602, Japan}

\affiliation{${}^6$Department of physics, Waseda University, Shinjyuku, Tokyo 169-8555, Japan}
\affiliation{${}^7$ Department of Physics, Kobe University, Kobe 657-8501, Japan}

\begin{abstract}

We investigate the relation between the $S$-matrix unitarity ($SS^\dagger=1$) and the renormalizability, in theories with negative norm states.
The relation has been confirmed in many theories, such as gauge theories, Einstein gravity and Lifshitz-type non-relativistic theories by analyzing the unitarity bound, which follows from the $S$-matrix unitarity and the norm positivity.
On the other hand, renormalizable theories with a higher derivative kinetic term do not necessarily satisfy the unitarity bound essentially because the unitarity bound does not hold due to the negative norm states.
In these theories, it is not clear if the $S$-matrix unitarity provides a nontrivial constraint related to the renormalizability. In this paper we introduce scalar field models with a higher derivative kinetic term
and analyze the $S$-matrix unitarity. 
We have positive results of the relation.

\end{abstract}

\maketitle

\section{Introduction}

Unitarity and renormalizability are fundamental principles in quantum field theories (QFTs). 
The renormalizability means that a theory can be described by a finite number of parameters. 
Fixing all of these parameters by experiments, we can make a nontrivial prediction.
On the other hand, the unitarity consists of two independent conditions: the $S$-matrix unitarity, i.e. $SS^\dagger=1$,
and the positivity of physical state norms (norm positivity).
The $S$-matrix unitarity guarantees a unique time evolution, whereas the positivity is to make the probability positive.
We emphasize that the $S$-matrix unitarity and the unitarity are carefully distinguished in this paper. 
Both unitarity and renormalizability are necessary to make the theory predictive, hence it is interesting to explore possible connections thereof.

\medskip
Indeed, there are a variety of known examples where the unitarity bound, that is derived from the $S$-matrix unitarity and the norm positivity, and the renormalizability imply the same constraints on interactions of the theory: Historical arguments include gauge theories and Einstein gravity~\cite{Llewellyn Smith:1973ey,Cornwall:1973tb,Cornwall:1974km, Berends:1974gk}. More recently, it was shown that this equivalence holds true in more generic QFTs such as Lifshitz-type non-relativistic scalar field theories~\cite{Fujimori:2015wda, Fujimori:2015mea}.
If this equivalence is universal, one may expect, e.g., that quantum gravity theories consistent with the unitarity bound are automatically renormalizable.
Since the unitarity bound significantly constrains the particle spectrum and interactions of quantum gravity theories (see, e.g.,~\cite{AH} for recent discussions), it would be helpful if we may use the unitarity bound as a probe of renormalizability in this way. We would like to explore this possibility by further investigating the possible equivalence.

\medskip
Let us here recall that the models~\cite{Llewellyn Smith:1973ey,Cornwall:1973tb,Cornwall:1974km, Berends:1974gk,Fujimori:2015wda, Fujimori:2015mea} mentioned above do not involve negative norm states and thus the unitarity bound directly follows from the $S$-matrix unitarity.
On the other hand, one may easily show by explicit calculations that renormalizable theories with higher derivative terms, such as the quadratic gravity, do not necessarily satisfy the unitarity bound. This is because the unitarity bound does not hold in these theories due to the negative norm states stemming from the higher derivatives.
Then, does it mean that the unitarity bound is not helpful anymore to investigate renormalizability?
To answer this question, we raise the following question: Does the $S$-matrix unitarity, without the positive norm assumption, provide nontrivial constraints?
In this paper we argue that the $S$-matrix unitarity indeed gives a nontrivial constraint even in theories with negative norm states. In particular, we demonstrate in a concrete model that the obtained constraint is identical to the one implied by the renormalizability.

\medskip
Among various models with negative norm states, we would like to focus on the quadratic gravity, whose action contains quadratic terms in the curvature tensor on top of the Einstein-Hilbert term. In the quadratic gravity, the graviton propagator is modified as $\sim p^{-4}$ in the UV region and thus the renormalizability is achieved at the cost of the norm positivity~\cite{Stelle:1976gc,Fradkin:1981iu}. Even though it is not clear yet how to deal with the negative norm states, the theory is predictive as long as the $S$-matrix is unitary, so that the quadratic gravity could offer an interesting approach to UV completion of gravity. In the present paper we consider a scalar field model with a modified propagator as a toy model for quadratic gravity and demonstrate that the $S$-matrix unitarity and the renormalizability lead to the identical constraint on this model. Our result provides a new evidence for the connection between the $S$-matrix unitarity and the renormalizability. It will also be useful for better understanding of quadratic gravity. In particular, our analysis implies that the unitarity bound is a useful probe for the renormalizability in theories without negative norm states, whereas the $S$-matrix unitarity has to be used instead if the theory contains negative norm states.

\medskip
The rest of this paper is organized as follows. 
In Sec.\ref{HDSFT}, we introduce a scalar field model with a higher derivative kinetic term. 
The scalar propagator in this model is modified as $\sim p^{-4}$ in the UV, so that we can think of it as a toy model for the quadratic gravity. 
We also elaborate on the renormalizability conditions in our model based on the references~\cite{Fujimori:2015wda, Fujimori:2015mea}. 
In Sec.\ref{OPUB}, we review the optical theorem and the unitarity bound. In particular, we argue that the lowest-order approximation of $SS^\dagger=1$ may be used to constrain the theory, just as the tree-level unitarity of high energy scattering constrains interactions in unitary theories. 
The lowest-order approximation of $SS^\dagger=1$ in the scalar field model is then investigated in Sec.\ref{SFMHOD}. 
We there find that the $S$-matrix unitarity and the renormalizability imply the identical constraints.
Summary and discussion are given in Sec~\ref{Summary}. Technical details are collected in Appendices.
 
\section{Higher derivative scalar field theory}\label{HDSFT}

In this section we introduce a scalar field model with a higher derivative kinetic term as a simplest example of theories with negative norm states.
Due to the fourth order derivative term in the kinetic term of the scalar field $\phi$, its propagator is modified as $\sim p^{-4}$ at UV and thus there appear negative norm (ghost) states. This model can then be thought of as a toy model for the quadratic gravity.
We also elaborate on the renormalizability conditions in this higher derivative theory based on the criterion introduced in Ref.~\cite{Fujimori:2015wda, Fujimori:2015mea}. 

\subsection{Quadratic action}

The quadratic action for the scalar field $\phi$ with the fourth order derivative in four dimensional spacetime is given~by
\begin{eqnarray}
\mathcal{S}_2=\int d^4x \left[-\frac{1}{2}\phi(\Box-m_{1}^2)(\Box-m_{2}^2)\phi \right],\label{HDSFT1}
\end{eqnarray}
where $\Box=\partial^{\mu}\partial_{\mu}$ and $m_{2}>m_{1}(>0)$.
The propagator of $\phi$ is given by $\displaystyle\frac{1}{(p^2+m_1^2)(p^2+m_2^2)}$, which has a mild UV behavior $\sim p^{-4}$. Also, it has poles at $p^2=-m_1^2$ and $p^2=-m_2^2$ representing two species of particles.
In order to construct the two sets of creation and annihilation operators,
it is better to rewrite the action into that of two scalars with the second order derivatives. 
It can be done by the following redefinition of the fields (a careful analysis using Lagrange multiplier is given in Appendix~\ref{2trans}),  
\begin{eqnarray}
\psi_{1}:=\frac{(\Box-m_{2}^2)\phi}{M}\hspace{1cm}\mbox{and}\hspace{1cm}
\psi_{2}:=\frac{(\Box-m_{1}^2)\phi}{M}\hspace{1cm} 
\left( \phi=\frac{1}{M}(\psi_2-\psi_1) \right), \label{phipsi}
\end{eqnarray}
where $M:=\sqrt{m_2^2-m_1^2}$.
Then, the action (\ref{HDSFT1})  becomes
\begin{eqnarray}
\mathcal{S}_2= \int d^4x \left[\frac{1}{2}\psi_{1}(\Box-m_{1}^2)\psi_{1}-\frac{1}{2}\psi_{2}(\Box-m_{2}^2)\psi_{2}\right]\,. \label{HDSFT2}
\end{eqnarray}
The kinetic term of $\psi_1$ has a positive sign, whereas that of $\psi_2$  a negative sign.
Thus, the one particle state of $\psi_{2}$ becomes a negative norm state (see Appendix\ref{Quantization}). 
The appearance of negative norm states is inevitable, if the original action has higher derivatives.
Note that $\phi$ is dimensionless, $[\phi]=0$, and $\psi_{i}$ has the canonical dimension, $[\psi_i]=1$, where the bracket means the mass dimension of parameter or field. We also introduce a canonical scalar field $\sigma$ by
\begin{eqnarray}
\mathcal{S}_{2\sigma}=\int d^4x \left[\frac{1}{2}\sigma(\Box-m_{\sigma}^2)\sigma \right] \label{HDSFTsigma}
\end{eqnarray}
to construct a nonrenormalizable coupling term later in section II.B.
This field $\sigma$ has dimension $[\sigma]=1$.

\subsection{Interaction terms}

We then introduce interaction terms. To discuss the relation between the tree-level $S$-matrix unitarity and the renormalizability, we consider both renormalizable and nonrenormalizable interactions, and clarify if they are consistent with $SS^\dagger=1$ or not in the perturbative expansion. Among various interactions, we focus on the marginal ones, where ``marginal" means that the mass dimension of the coupling constant is naught.
It is first because other cases can easily be inferred from the results in the marginal cases. Moreover, if the theory contains a field with a non-positive dimension (our $\phi$ field has the vanishing dimension), the renormalizability condition is not as simple as ordinary QFTs~\cite{Fujimori:2015wda,Fujimori:2015mea}. As we explain shortly, the renormalizability condition of marginal interactions does not simply follow from the ordinary power-counting, but rather requires an additional condition in our setup. 
Therefore, we expect to see similar nontrivial conditions for the $S$-matrix unitarity, that is the main objective of this paper. We also concentrate on the tree-level four-point scattering because it is simple, but provides enough interesting information for our purpose. Thus, we will consider marginal four-point vertex operators.

\medskip 
Before introducing interaction terms, 
we briefly summarize the results in Ref.~\cite{Fujimori:2015wda,Fujimori:2015mea}. Let us start from the general discussion of the following $n$-th order interaction in $4$ dimensional spacetime,
\begin{eqnarray}
S_{int} = \lambda \int d^{4} x \, (\partial_x^{a_1} \phi_1) \, (\partial_x^{a_2} \phi_2) \cdots (\partial_x^{a_n} \phi_n), 
\label{eq:vertex}
\end{eqnarray}
where $a_1,\cdots, a_n$ are non-negative integers. 
Fields $\phi_i$ with different indices $i$ can be either the same or different.
Each $\partial_x$ denotes spacetime derivatives $\partial_\mu~(\mu=0,\cdots, 3)$, whose indices are contracted in a Lorentz invariant way.
In the interaction action (\ref{eq:vertex}), the dimension of the coupling constant $\lambda$ reads
\begin{eqnarray}
[\lambda] = 4 - \sum_{l=1}^n ( a_l + [\phi_l] ).
\label{eqlam}
\end{eqnarray}
The usual power-counting renormalizability (PCR) condition requires that the mass dimension of the coupling constant is non-negative, $[\lambda]\geq0$. 
However, an infinite number of interaction terms satisfy this condition if there exists a field with a non-positive dimension $[\phi_l]\le0$. 
In other words, an infinite number of parameters are required, and thus the theory fails to be renormalizable.
To make the number of parameters to be finite, we introduce the extended PCR conditions given by
\begin{eqnarray}
\sum_{~l=1~}^n (a_l + [\phi_l]) &\le& 4, \\
\sum (a_l + [\phi_l]) &<&  4
\quad
\mbox{for an arbitrary partial sum}.
\label{eq:ePCR}
\end{eqnarray}
The former is nothing but the usual PCR condition, while the latter additional conditions mean that the dimension of $\partial_x^{a_{i_1}} \phi_{a_{i_1}} \cdots \partial_x^{a_{i_k}} \phi_{a_{i_k}}~(k\le n-1)$ should be less than the spacetime dimension $4$.
In particular we can see that the second conditions provide nontrivial constraints only on marginal interactions in our setup, which contains a field $\phi$ with the vanishing dimension, but does not contain fields with negative dimensions.
Based on the extended PCR conditions, we introduce two marginal interaction terms. 
One is renormalizable, while the other is not.

\subsubsection{Renormalizable interaction terms} 

\begin{figure}[t]
\begin{center}
\includegraphics[width=105mm]{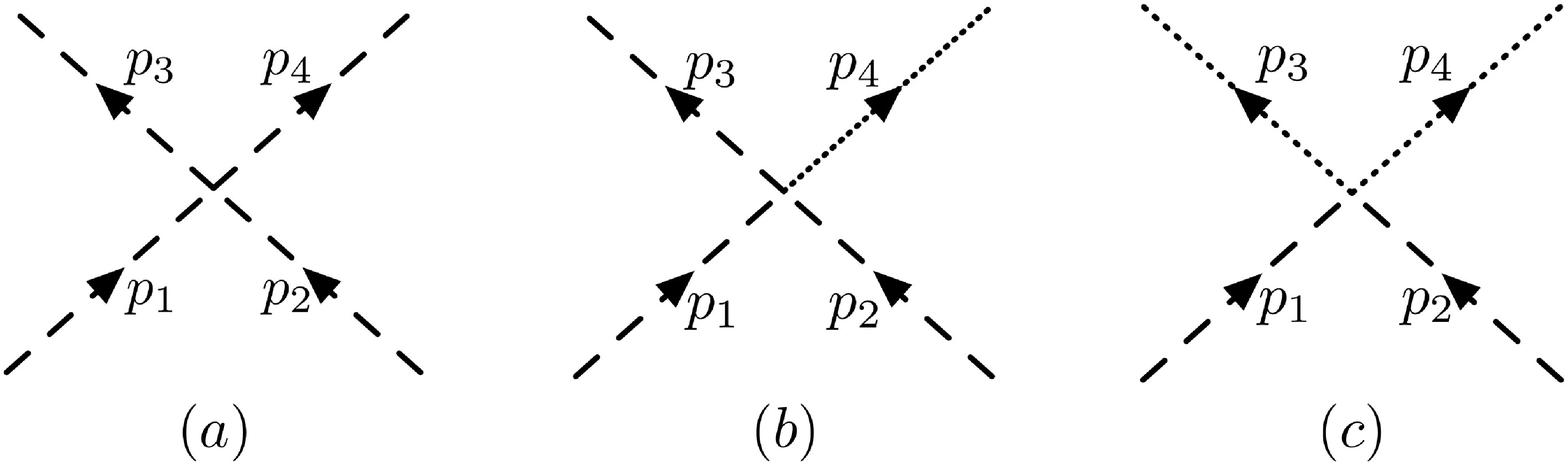}
\vspace{0mm}
\vskip-\lastskip
\caption{The scalar four-point vertex functions (a)$\psi_{1}\psi_{1}\psi_{1}\psi_{1}$, (b)$\psi_{1}\psi_{1}\psi_{1}\psi_{2}$, and (c)$\psi_{1}\psi_{1}\psi_{2}\psi_{2}$.
The broken and dotted lines stand for the scalar fields $\psi_{1}$ and $\psi_{2}$, respectively.}
\label{fig:4pointvertex11XX}
\end{center}
\end{figure}

We consider the simplest renormalizable marginal interaction term. 
\begin{eqnarray}
S_{ren} = \lambda \int d^4x \, \left\{( \partial_\mu \phi)^2\right\}^2\,.
\label{4th}
\end{eqnarray}
This interaction has $a_1=a_2=a_3=a_4=1$, and satisfies the extended PCR conditions (8). For the later calculations of the scattering amplitudes, it is better to rewrite it in terms of $\psi_1$ and $\psi_2$ defined in eqs.(\ref{phipsi}) (the reformulation using the Lagrange multiplier in Appendix~\ref{2trans} gives the same result),
\begin{eqnarray}
&&S_{ren} = \alpha \int d^4x \, \left\{\big( \partial_\mu (\psi_2-\psi_1)\big)^2\right\}^2,
\label{4thpsi} \\
&&\alpha :=  \frac{\lambda}{M^4}.
\end{eqnarray}
This gives the four-point interaction terms for $\psi_1$ and $\psi_2$ with derivatives, 
and they have the same coupling constant $\alpha$, with $[\alpha]=-4$. 
Although $[\alpha]<0$ does not satisfy the PCR condition, the theory turns out to be renormalizable because the common coupling $\alpha$ brings about a cancelation among the $\psi_{1}$ loop and the $\psi_{2}$ loop.
In the next section, we use the vertex functions (a)$\psi_{1}\psi_{1}\psi_{1}\psi_{1}$, (b)$\psi_{1}\psi_{1}\psi_{1}\psi_{2}$, and (c)$\psi_{1}\psi_{1}\psi_{2}\psi_{2}$ given by (see FIG.\ref{fig:4pointvertex11XX})
\vspace{0mm}
\begin{equation}
\begin{split}
(a)&=8i\alpha\left[\left(p_{1}\cdot p_{2}\right)\left(p_{3}\cdot p_{4}\right)+\left(p_{1}\cdot p_{3}\right)\left(p_{2}\cdot p_{4}\right)+\left(p_{1}\cdot p_{4}\right)\left(p_{2}\cdot p_{3}\right)\right],\\
(b)&=-8i\alpha\left[\left(p_{1}\cdot p_{2}\right)\left(p_{3}\cdot p_{4}\right)+\left(p_{1}\cdot p_{3}\right)\left(p_{2}\cdot p_{4}\right)+\left(p_{1}\cdot p_{4}\right)\left(p_{2}\cdot p_{3}\right)\right],\\
(c)&=8i\alpha\left[\left(p_{1}\cdot p_{2}\right)\left(p_{3}\cdot p_{4}\right)+\left(p_{1}\cdot p_{3}\right)\left(p_{2}\cdot p_{4}\right)+\left(p_{1}\cdot p_{4}\right)\left(p_{2}\cdot p_{3}\right)\right],
\label{4pointvertex11XXa}
\end{split}
\end{equation}
where $p_{i} (i=1, 2, 3, 4)$ are the 4-momenta of scalar fields.

\subsubsection{Non-renormalizable interaction terms}
\begin{figure}[t]
\begin{center}
\includegraphics[width=70mm]{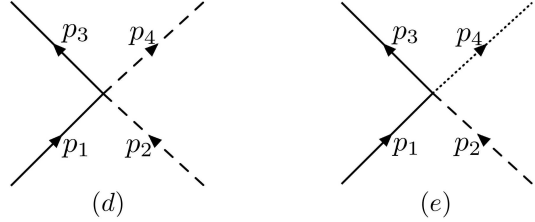}
\vspace{0mm}
\vskip-\lastskip
\caption{The scalar four-point vertex functions (d)$\sigma\psi_{1}\sigma\psi_{1}$ and (e)$\sigma\psi_{1}\sigma\psi_{2}$.
The solid, broken and dotted lines stand for the scalar fields $\sigma$, $\psi_{1}$ and $\psi_{2}$, respectively.}
\label{fig:4pointvertexS1SX}
\end{center}
\end{figure}

We desire to consider the simplest examples of non-renormalizable marginal interaction terms, that would be
\begin{eqnarray}
\int d^4 x \phi(\partial_{\mu}\phi)^2(\Box\phi)
\hspace{1cm}\mbox{or}\hspace{1cm}
 \int d^4 x \phi^2(\Box\phi)^2
 \label{4DNSI}.
\end{eqnarray}
However, it turns out that we have to analyze six-point scattering amplitudes to check their consistency with the $S$-matrix unitarity $SS^\dagger=1$, essentially because these interactions contain the $\Box$ operator (as shown in Appendix~\ref{Box}).
They are not suitable for our analysis based on four-point scattering amplitudes.
Notice here that any marginal fourth order interaction with four derivatives can be reformulated into the three interaction terms in (\ref{4th}) and (\ref{4DNSI}) by partial integrals.
We then introduce another scalar field $\sigma$ defined in (\ref{HDSFTsigma}) and consider the following interaction as a simple example of non-renormalizable marginal interactions:
\begin{eqnarray}
S_{non}=\lambda' \int d^4 x \phi^2(\partial_{\mu}\sigma)^2\label{4th-non},
\end{eqnarray}
where $[(\partial_{\mu}\sigma)^2]=4$, and thus this coupling term does not satisfy the extended PCR conditions (8).
This interaction term can be expressed with $\psi_1$ and $\psi_2$ as 
\begin{eqnarray}
&&S_{non}=\alpha' \int d^4 x (\psi_2-\psi_1)^2(\partial_{\mu}\sigma)^2\label{4thpsi-non}, \\
&&\alpha':= \frac{\lambda'}{M^2}.
\end{eqnarray}
In the next section we use the vertex functions $\sigma\psi_{1}\sigma\psi_{1}$ and $\sigma\psi_{1}\sigma\psi_{2}$ (see FIG.\ref{fig:4pointvertexS1SX}) given by
\vspace{0mm}
\\
\begin{equation}
\begin{split}
(d)&=-4i\alpha'\left(p_{1}\cdot p_{3}\right),\\
(e)&=4i\alpha'\left(p_{1}\cdot p_{3}\right),
\label{4pointvertexS1SX}
\end{split}
\end{equation}
where $p_{i} (i=1, 3)$ are the 4-momenta of scalar fields.

\section{$S$-matrix unitarity and renormalizability}
\label{Secud}

We then investigate the relation between the renormalizability and the $S$-matrix unitarity for high energy scattering. In theories without negative norm states, the $S$-matrix unitarity implies the unitarity bound. We know that the unitarity bound gives the identical constraints as the renormalizability in various theories.
On the other hand, if there exist negative norm states, the unitarity bound does not follow from the $S$-matrix unitarity anymore. However, in this section, we argue that the optical theorem (which is based on the $S$-matrix unitarity, but does not rely on the norm positivity) provides nontrivial consistency conditions even in theories with negative norm states. Interestingly, we find that the obtained constraint is identical to the one required by renormalizability. 

\medskip
In Sec.\ref{OPUB}, we firstly see the relation among the $S$-matrix unitarity, the optical theorem and the unitarity bound, by carefully reviewing the derivation of the unitarity bound and locating the step in which the positivity of all norms is required. 
There, we will see that the optical theorem may be used to constrain theories with negative norms. In Sec.\ref{SFMHOD}, then, we concretely investigate the relation between the renormalizability and the optical theorem, 
in the simple scalar field model introduced in Sec.\ref{HDSFT}.
In particular, we will see that the (non)renormalizable interaction term is (in)consistent with the optical theorem in the UV limit.

\subsection{Optical theorem and unitarity bound}\label{OPUB}

We begin with the $S$-matrix unitarity,
\begin{eqnarray}
SS^\dagger =  1.
\label{US}
\end{eqnarray}
Decomposing the S-matrix as $S =  1 + i T$,
we can rewrite the above equation in 
\begin{eqnarray}
-i(T - T^{\dagger})=TT^{\dagger}.
\label{T1}
\end{eqnarray}
Let us also suppose that we can take a complete orthonormal basis of the Hilbert space $\{ |{X} \rangle \}$ as
\begin{eqnarray}
\sum_{X}|X\rangle C_X \langle X|=1 \hspace{1cm} \mbox{with} \hspace{1cm}
 \langle X|Y\rangle=C_X^{-1}\delta_{XY},
\label{defCX}
\end{eqnarray}
where we assumed that $\langle X|X\rangle\neq 0$, but it can be either positive or negative.
Note that if the theory contains a negative norm state, its normalization factor $C_X$ is inevitably negative, because the sign of $C_X$ cannot be flipped by the normalization of the state $|X\rangle$.
We then would like to reformulate the relation~\eqref{T1} in terms of the covariant scattering amplitude ${\cal{M}}{\left(i \to f\right)}$ defined as
\begin{eqnarray}
\langle f |T| i \rangle = \delta^{4} \left( {p}_i - {p}_f \right) {\cal{M}} \left( i \to f \right),
\label{defM}
\end{eqnarray}
By substituting eq.(\ref{defM}) into eq.(\ref{T1}), we arrive at the cutting rule in terms of covariant amplitudes,
\begin{eqnarray}
-i[{\cal{M}}\left(i\to f\right)-{\cal{M}}\left(f\to i \right)^{\ast}] = \sum_{X}C_X\delta^{4}  \left( {p} - {p}_X \right){\cal{M}}\left(i\to X\right){\cal{M}}\left(f\to X\right)^{\ast},
\end{eqnarray}
where ${ p}(={p}_{i}={p}_{f})$ is the total energy-momentum. 
By considering the forward scattering ($| i \rangle=| f \rangle$), the relation is reduced to the optical theorem,
\begin{eqnarray}
2{\rm{Im}}{\cal{M}}\left(i\to i\right)=\sum_{X} C_X \delta^{4}  \left( {p} - {p}_X \right)|{\cal{M}}\left(i\to X\right)|^2,
\label{op}
\end{eqnarray}
where the summation is over all possible intermediate on-shell states. In particular, the optical theorem implies 
\begin{eqnarray}
|{\cal{M}}\left(i\to i\right)|\ge|{\rm{Im}}{\cal{M}}\left(i\to i\right)|
=\frac{1}{2}\left|\sum_{X} C_X \delta^{4}  \left( {p} - {p}_X \right)|{\cal{M}}\left(i\to X\right)|^2\right|. \label{ImM}
\end{eqnarray}
The discussion so far is applicable even if the theory contains negative norm states.

\medskip
If the theory does not contain negative norm states, the inequality~\eqref{ImM} implies the unitarity bound. To explain it, let us consider theories without negative norm states and set $C_X=1$ by taking the normal basis.
In this case, the right hand side of \eqref{ImM} is bounded as
\begin{align}
\text{r.h.s. of \eqref{ImM}}&=\frac{1}{2}\sum_{X}\delta^{4}  \left( {p} - {p}_X \right)|{\cal{M}}\left(i\to X\right)|^2\geq\frac{1}{2}|{\cal{M}}\left(i\to j\right)|^2
\end{align}
for an arbitrary on-shell intermediate state $j$. By setting $j=i$, we obtain an inequality, 
\begin{eqnarray}
|{\cal{M}}\left(i\to i\right)|\le const, \label{UB1}
\end{eqnarray}
which further implies that the left hand side of (\ref{ImM}) is bounded from above by a constant. By combining eqs.~(\ref{ImM})-(\ref{UB1}), we arrive at the following bound on scattering amplitudes with an arbitrary on-shell state $j$:
\begin{eqnarray}
|{\cal{M}}\left(i\to j\right)|\le const. \label{UB}
\end{eqnarray}
This is the unitarity bound.\footnote{To determine the precise value of the upper bound (which was simply denoted by ``const."), we need to carefully evaluate the normalization factor originated from the delta function in (\ref{defM}). See, e.g,~\cite{Fujimori:2015mea} for details.} In particular, this bound for high energy scattering at the tree-level (that is, the lowest order in coupling constants) is called the tree-level unitarity condition.

\medskip
It has been known that, the unitarity bound (\ref{UB}), especially the tree-level unitarity, gives the identical condition as the renormalizability in various theories without negative norm states.
A natural question will be if the weaker bound~\eqref{ImM}, which holds true even in theories with negative norm states, may provide nontrivial constraints useful to analyze the UV properties such as renormalizability.
In the next subsection, we use the scalar field model introduced in Sec.\ref{HDSFT} to demonstrate that it is indeed the case. In particular, we show that tree-level high energy scattering induced by the (non)renormalizable interaction is (in)consistent with the bound~\eqref{ImM} implied by the optical theorem.

\subsection{Weaker bound for amplitudes in higher derivative scalar models}\label{SFMHOD}

We finally come to our main question of whether the weaker unitarity bound (\ref{ImM}) is satisfied for amplitudes in higher derivative scalar models of Sec. {\ref{HDSFT}}. 
In two ways of quantization for $\psi_2$ (see Appendix\ref{Quantization}), we choose the one that negative norm states exist.%
\footnote{
Choosing the other quantization creating negative-energy particles, no negative norm state appears. 
The cancelation in (\ref{op3}), that we will explain, never occurs, and thus the theory does not possess the perturbative $S$-matirix unitary. 
The relation between the renormalizability and the $S$-matrix unitarity suggests that 
the perturbative renormalizability property is never better in such quantization with higher order derivatives.
}
For the reason mentioned in Sec. \ref{OPUB}, we consider two-two scattering 
\begin{eqnarray} 
{p}_1 + {p}_2  \to {p}_3 + {p}_4 \label{1}\,.
\end{eqnarray} 
It is customary to work in the center of mass frame, in which the three momenta satisfy 
\begin{eqnarray} 
{\bf p}_2 = - {\bf p}_1, \hspace{3mm}{\bf p}_4 = - {\bf p}_3  .\label{2}
\end{eqnarray} 
We denote the scattering angle by $\theta$. 
We consider two-two scattering of $\psi_{1}, \psi_{2}, \sigma,$ 
\begin{subequations}
\begin{align}
\psi_1({\bf p}_1)\psi_1({\bf p}_2) &\to\psi_1({\bf p}_3)\psi_1({\bf p}_4)\label{3a} \,,
\\[1mm]
\psi_1({\bf p}_1)\psi_1({\bf p}_2) &\to\psi_1({\bf p}_3)\psi_2({\bf p}_4)\label{3b} \,,
\\[1mm]
\psi_1({\bf p}_1)\psi_1({\bf p}_2) &\to \psi_2({\bf p}_3)\psi_2({\bf p}_4) \label{3c} \,,
\\[1mm]
\sigma({\bf p}_1)\psi_1({\bf p}_2) &\to \sigma({\bf p}_3)\psi_1({\bf p}_4)\label{3d}\,,
\\[1mm]
\sigma({\bf p}_1)\psi_1({\bf p}_2) &\to \sigma({\bf p}_3)\psi_2({\bf p}_4) \label{3e}\,.
\end{align} 
\end{subequations}
Note that (\ref{3b}) and (\ref{3e}) contain a single negative norm particle $\psi_{2}$. The amplitudes are denoted by 
\begin{eqnarray} 
{\cal M}\big(\psi_1({\bf p}_1)\psi_1({\bf p}_2)\to\psi_1({\bf p}_3)\psi_1({\bf p}_4)\big) \label{4a} 
\end{eqnarray} 
for the process (\ref{3a}), and similarly for the other four. The calculation of the amplitudes is somewhat technical and is given in Appendix \ref{calSA}.

\medskip
We first give the results for the elastic amplitudes for positive norm particles given in \eqref{E7} and \eqref{E15} as
    \begin{eqnarray}
&&{\cal M}\big(\psi_1({\bf p}_1)\psi_1({\bf p}_2)\to\psi_1({\bf p}_3)\psi_1({\bf p}_4)\big) =8\alpha\big( (6+ 2\cos^2\theta)|{\bf p}_1|^4+8|{\bf p}_1|^2m_1^2+3m_1^4\big), \label{SA1} \\
&&
{\cal M}\big(\sigma({\bf p}_1)\psi_1({\bf p}_2)\to\sigma({\bf p}_3)\psi_1({\bf p}_4)\big) =-4\alpha'\big( (1-\cos\theta)|{\bf p}_1|^2+m_\sigma^2\big),\label{SA2}
\end{eqnarray} 
where the former is the case of renormalizable interaction, and the latter the non-renormalizable one. We immediately note that both amplitudes diverge in the limit of $|{\bf p}_1| \to \infty$. Hence the unitarity bound (\ref{UB}) is apparently violated for both renormalizable and non-renormalizable marginal interactions, (\ref{4th}) and (\ref{4DNSI}) respectively of Sec \ref{HDSFT}. This result is not unexpected because our model contains negative norm states. 

\medskip
We then discuss if these two amplitudes are consistent with the weaker bound~\eqref{ImM} implied by the optical theorem. 
Similarly to the tree-level unitarity argument, we work in the lowest order approximation in the coupling constants. 
In this approximation, only the intermediate two-particle states contribute to the right hand side of the relation~\eqref{ImM}: 
\begin{eqnarray}
&&\Big|{\cal M}\left(\Psi_1({\bf{p}}_{1}),\Psi_2({\bf{p}}_{2}) \to \Psi_1({\bf{p}}_{1}),\Psi_2({\bf{p}}_{2})\right)\Big| \nonumber \\
&&\qquad 
\ge  \Big|\rm{Im}{\cal M}\left(\Psi_1({\bf{p}}_{1}),\Psi_2({\bf{p}}_{2}) \to \Psi_1({\bf{p}}_{1}),\Psi_2({\bf{p}}_{2})\right) \Big|\nonumber \\
&&\qquad
= \frac{1}{2}\left| \sum_{{i}{j}}\int \frac{d^3{\bf p}_i}{ 2E_i} \frac{d^3{\bf p}_j}{ 2E_j} \delta^4 (p_1+p_2-p_i-p_j) (-1)^{n_{ij}} \Big| {\cal M} \left( {\Psi_1(\bf{p}}_{1}),\Psi_2({\bf{p}}_{2})\to\Psi_i({\bf{p}}_{i}),\Psi_j({\bf{p}}_{j})\right) \Big|^2\right| \label{op2},
\end{eqnarray}
where $\Psi_i({\bf p}_i)$ means a one-particle state of $\psi_1$, $\psi_2$ or $\sigma$  with the three dimensional momentum ${\bf p}_i$, $n_{ij}$ is the number of negative norm particles $\psi_2$ in the state $|\Psi_i({\bf{p}}_{i}),\Psi_j({\bf{p}}_{j})\rangle$ and we take all possible two-particle states in the sum with respect to $i$ and $j$.
$E_i$ is the energy of the intermediate on-shell particles $i$, and similarly for other on-shell momenta.
The measure of integral is fixed by the inner product (\ref{innerP}) in Appendix~\ref{Quantization}.
In the center of mass frame (${\bf p}_1=-{\bf p}_2$, ${\bf p}_i=-{\bf p}_j$), the last line of eq.(\ref{op2}) can be written as (see Appendix~\ref{Doop3})
\begin{eqnarray}
\propto\left|\sum_{ij} (-1)^{n_{ij}} \frac{ |\hat{{\bf p}}_i|}{(E_1+E_2)} \int_0^\pi \sin \theta \Big| {\cal M} \left( \Psi_1({\bf{p}}_{1}),\Psi_2(-{\bf{p}}_{1})\to\Psi_i(\hat {\bf{p}}_{i}),\Psi_j( -\hat{\bf{p}}_{i})\right) \Big|^2 d\theta \right|   \label{op3},
\end{eqnarray}
where $\theta$ is the angle between ${\bf p}_1$ and ${\bf p}_i$. Also, $\hat{\bf p}_i$ is the on-shell momentum satisfying the energy-momentum conservation, so that its absolute value $|\hat{\bf p}_i|$ is a function of $|{\bf p}_1|$ and $\theta$. 
The proportionality constant is an unimportant numerical factor.

\medskip
We can now compare the energy dependence of the both sides in the inequality (\ref{op2}) in the high energy limit.
Since we saw in eq.(\ref{SA1}) that ${\cal M}\big(\psi_1({\bf p}_1)\psi_1({\bf p}_2)\to\psi_1({\bf p}_3)\psi_1({\bf p}_4)\big)$ behaves as $|{\bf p}_1|^4$ in the UV limit, 
the naive comparison of energy dependence in each term seems to imply violation of the inequality~\eqref{op2} even for the renormalizable interaction. 
However, we will find that negative contributions in the sum of the optical theorem (\ref{op2}) brings about a cancellation, which restores the optical theorem for the renormalizable interaction.
On the other hand, we will find that the cancellation for the nonrenormalizable case is not enough to restore the inequality~\eqref{op2}.
The argument below is based on the calculation of each scattering amplitude ${\cal M} \left( \Psi_1({\bf{p}}_{1}),\Psi_2(-{\bf{p}}_{1})\to\Psi_i(\hat {\bf{p}}_{i}),\Psi_j( -\hat{\bf{p}}_{i})\right)$ in Appendix \ref{calSA}.

\subsubsection{Renormalizable interaction term}

We first consider the renormalizabe interaction (12) and  investigate the inequality (\ref{op2}). The calculation is simple if we take the center-of-mass initial state with two $\psi_1$, $|i\rangle=|\psi_1,\psi_1 \rangle$.
The left hand side of (\ref{op2}) is obtained by setting $\cos \theta =1$ in the amplitude (\ref{SA1}), 
\begin{eqnarray}
&&{\cal M}\big(\psi_1({\bf p}_1)\psi_1(-{\bf p}_1)\to\psi_1({\bf p}_1)\psi_1(-{\bf p}_1)\big) =8\alpha\big( 8|{\bf p}_1|^4+8|{\bf p}_1|^2m_1^2+3m_1^4\big) = \alpha {\cal O}\left(|{\bf p}_1|^4\right).
\end{eqnarray} 
The right hand side of inequality (\ref{op2}) is the sum of (\ref{1111}), (\ref{1112}) and (\ref{1122}). 
Here, we have to multiply (\ref{1112}) by $2$ because there are two cases, $(\Psi_i,\Psi_j)=(\psi_1,\psi_2),(\psi_2,\psi_1)$. 
Then, it can be easily seen that the cancelation occurs in the leading order and the next leading order. 
As a result, the right hand side of the inequality (\ref{op2}) is 
\begin{eqnarray} 
&&64 \alpha^2\frac{m_1^4|{\bf p}_1|^5}{E_1+E_2}\int_0^\pi  {\cal O}\left( \left( \frac{m_1}{|{\bf p}_1|}\right)^0 \right)     d \theta= \alpha^2 {\cal O} \left(|{\bf p}_1|^4\right).
\end{eqnarray} 
Hence, the energy dependence of both sides of the inequality (\ref{op2}) at high energy is the same. 
Therefore, the inequality (\ref{op2}) is satisfied at high energy, provided we take the coupling constant $\alpha$ sufficiently small.

\subsubsection{Nonrenormalizable interaction term}

Next, we will see that the inequality (\ref{op2}) does not hold true for nonrenormalizable interaction term (\ref{4th-non}).
The left hand side of $(\ref{op2})$ is the amplitude for the forward scattering and it can be obtained by setting $\cos \theta=1$ in eq.(\ref{SA2}),
\begin{eqnarray}
{\cal M}\big(\sigma({\bf p}_1)\psi_1(-{\bf p}_1)\to\sigma({\bf p}_1)\psi_1(-{\bf p}_1)\big) =4\alpha' m_\sigma^2.
\end{eqnarray} 
Note that the leading order term in the high energy limit, that is proportional to $\left|{\bf p}_1\right|^2$, disappears and that this approaches to a constant in the high energy limit. 
The right hand side is the sum of (\ref{s1s1}) and (\ref{s1s2}). 
The leading order of the sum is canceled, but the next leading order is not. 
The exact form of the sum is 
\begin{eqnarray} 
&&16 {\alpha'}^2\frac{|{\bf p}_1|^5}{E_1+E_2}\int_0^\pi \sin\theta  \left[ (1-\cos\theta)^2 \frac{3M^2}{4|{\bf p}_1|^2}+ {\cal O}\left(\left(\frac{m_1}{|{\bf p}_1|}\right)^4\right) \right]     d \theta = \alpha'^2 {\cal O}(|{\bf p}_1|^2).
\end{eqnarray} 
This means that the right hand side diverges in the high energy limit.
Therefore, the inequality (\ref{op2}) is not satisfied in the high energy limit, i.e.  
the optical theorem is violated at some energy.

\section{Summary} \label{Summary}

We have studied the relation between the $S$-matrix unitarity $SS^\dagger =1$ and the renormalizability in scalar field models with a higher derivative kinetic term, which gives negative norm states. 
Two cases, a theory with a renormalizable interaction and another with a non-renormalizable interaction, have been investigated.
The tree-level $S$-matrix unitarity is satisfied in the former, while it is violated in the latter. 
They are evidences that the $S$-matrix unitarity and the renormalizability are related to each other even in theories with negative norm states.

\medskip
Combining our result with the past analyses of the relation between the unitarity bound and the renormalizability without negative norm states~\cite{Llewellyn Smith:1973ey,Cornwall:1973tb,Cornwall:1974km, Berends:1974gk,Fujimori:2015wda, Fujimori:2015mea}, 
the relation between the $S$-matrix unitarity and the renormalizability is expected to hold true in generic QFTs, no matter if there are negative norm states or not.
Our results therefore support an expectation that the unitarity bound is useful to investigate the renormalizability of theories with the norm positivity.
We believe that it is powerful especially when applied to quantum gravity theories.

\medskip
The relation to the renormalizability of $R_{\mu\nu}^2$ gravity is a particularly interesting objective. 
$R_{\mu\nu}^2$ gravity is known to be renormalizable. 
The matter scattering of canonical fields (i.e. non-ghost fields) with graviton propagator is confirmed to satisfy the unitarity bound \cite{Abe:2017abx}. 
It is interesting to see the optical theorem for graviton scattering.
The counter terms in the renormalization can be combined due to the symmetry (i.e. the general covariance), 
and thus the number of required parameters of counter terms is finite. 
Our result suggests that the $S$-matrix unitarity also requires the symmetry. 
It would be possible to be seen in graviton scattering.
From  this point of view, it is also very interesting to analyze the graviton loops to the scalar potential in $R^{2}_{\mu\nu}$ gravity theory.
Actually, in Einstein gravity, we have known that gravitational corrections give rise to nonrenormalizable divergent terms \cite{Smolin:1979ca, Bhattacharjee:2012my, Abe:2016irv}.
It is suggested that $R^{2}_{\mu\nu}$ gravity is a useful attempt at this problem \cite{Salvio:2014soa, Abe:2016hdj}.

\section*{Acknowledgements}

The authors would like to thank Y.-T.\,Huang for valuable discussions.
T.\,I. thanks Yoonbai Kim and S. Kawai for the kind hospitality during his one and a half year stay at Sungkunkwan University, where he benefitted much from the discussions with the theory people and the students. 
T.\,K. thanks K.\,Maeda, S.\,Yamada and S.\,Miyashita for valuable discussions. 
K.\,I. is supported by a JSPS Grants-in-Aid for Young Scientists (B) (No.\,17K14281) and for Scientific Research (A)(No.\,17H01091).
T.\,K. is supported by a Waseda U. Grant-in-Aid for Special Research Projects (2017K-233).
T.\,N.  is in part supported by JSPS KAKENHI Grant Numbers JP17H02894 and JP18K13539, and MEXT KAKENHI Grant Number JP18H04352.
The authors wish to thank T.\,Hatsuda of Riken, Y.\,Tsuboi of Chuo U for their kind hospitality.

\appendix

\section{Decomposition to two scalar modes}\label{2trans}

In Sec.\ref{HDSFT}, we rewrite the action (\ref{HDSFT1}) into the action (\ref{HDSFT2}) by the redefinition of fields (\ref{phipsi}).
One may wonder whether the transformation involving derivatives is allowed.  
We show in this appendix that this redefinition is consistent with the reformulation by introducing a Lagrange multiplier and an auxiliary field.

We introduce the auxiliary field $\chi$ as 
\begin{eqnarray}
\chi = \Box \phi.\label{chi}
\end{eqnarray}
It can be done by introducing the Lagrange multiplier $\lambda$ and the action (\ref{HDSFT1}) can be expressed as
\begin{eqnarray}
\mathcal{S}_2=\int d^4x \left[-\frac{1}{2}(\chi-m_{1}^2\phi)(\chi-m_{2}^2\phi) -\lambda (\chi-\Box \phi)\right].\label{HDSFT3}
\end{eqnarray}
We can easily check that the variation of the above action with respect to $\lambda$ gives eq.(\ref{chi}) and that, 
substituting eq.(\ref{chi}), the above action is reduced to the original action (\ref{HDSFT1}). 
In the action (\ref{HDSFT3}), $\chi$ does not have the kinetic term, and thus the variation with respect to $\chi$ give a 
constraint equation
\begin{eqnarray}
\chi=\frac{1}{2}\left(m_1^2+m_2^2\right)\phi - \lambda.
\end{eqnarray}
Substituting it into the action (\ref{HDSFT3}), we have
\begin{eqnarray}
\mathcal{S}_2=\int d^4x \left[\lambda \Box \phi + \frac{1}{2}\lambda^2 -\frac{1}{2}\left(m_1^2+m_2^2\right)\lambda \phi+\frac{1}{8}\left(m_1^2-m_2^2\right)^2 \phi^2\right].\label{HDSFT4}
\end{eqnarray}
This action can be diagonalized by the following field redefinition
\begin{eqnarray}
\lambda= -\frac{M}{2} (\psi_1+\psi_2) \hspace{1cm} \mbox{and}\hspace{1cm}
\phi=\frac{(\psi_2-\psi_1)}{M},\label{lambdaphi2} 
\end{eqnarray}
where $M=\sqrt{m_2^2-m_1^2}$. 
We can see that the form of $\phi$ is the same as that in eq.(\ref{phipsi}).
Then, the action (\ref{HDSFT4}) becomes
\begin{eqnarray}
\mathcal{S}_2= \int d^4x \left[\frac{1}{2}\psi_{1}(\Box-m_{1}^2)\psi_{1}-\frac{1}{2}\psi_{2}(\Box-m_{2}^2)\psi_{2}\right], \label{HDSFT5}
\end{eqnarray}
which is the same as action (\ref{HDSFT2}). 

For interaction term ${\mathcal L}_{int} (\partial, \phi)$, 
we can substitute eq.(\ref{lambdaphi2}). 
The reason is as follows. 
When we introduce the auxiliary field $\chi$, we can leave $\phi$ in the interaction term untouched. 
Then, the elimination of $\chi$ can be done in the same way and we do the field redefinition (\ref{lambdaphi2}).
This is nothing but substituting eq.(\ref{lambdaphi2}).

\section{Interaction with $\Box$ operator} \label{Box}

In this paper, we discuss the relation between  the renormalizability and the $S$-matrix unitarity. 
However, in the latter analysis, 
if the interaction term has $\Box$ operators attached to external lines, 
the amplitude is suppressed. 
Therefore, in this order, we can not see the relation.
If we go to the higher-point amplitude, we would see the correspondence, 
but the calculation of higher point amplitude is involved. 
We briefly show why $\Box$ operators make the discussion different.

We naively expect that $\Box$ operator gives the ${\cal O}(E^2)$ contribution. 
However, if the $\Box$ operator acts on an external line, it becomes $m^2(={\cal O}(E^0))$ by the on-shell condition,
that is, the UV behavior becomes mild.  
It is also true even in the canonical theory (without negative norm states). 
For instance, we consider the following action for a canonical scalar field $\zeta$
\begin{eqnarray}
{\mathcal S} = \int d^4x \left[ \frac{1}{2} \zeta \left( \Box-m^2\right) \zeta + \alpha_\zeta \zeta^3 \Box \zeta \right].
\end{eqnarray} 
Based on the power-counting argument, the interaction term is not renormalizable. 
Meanwhile, the tree-level four-point scattering amplitude is constant, and thus it satisfies the unitarity bound. 
To check the violation of the unitarity bound, we have to consider a tree diagram that has internal lines where $\Box$ operator is assigned. 
If we consider the six-point amplitude, an internal line appears, and thus it is expected to violate the unitarity bound. 

The situation is the same in the case with higher derivative that we analyze in this paper.  
We have seven kinds of four-point marginal vertices. 
Nevertheless, they are related to the integration by parts, and actually all of them can be expressed by the following three vertex functions,
\begin{eqnarray}
(\Box^2 \phi)\phi^3, \hspace{1cm} (\Box \phi)^2 \phi^2, \hspace{1cm} \left((\partial_\mu \phi)^2\right)^2 \label{4pt}.
\end{eqnarray} 
The last term is considered in this paper as a renormalizable vertex function. 
The other two are nonrenormalizable vertex terms, but because of $\Box$ operator,  
we have to go the six-point scattering amplitude to see the violation of the $S$-matrix unitarity.  

Another possible way to see the correspondence for nonrenormalizable operator would be considering a three point marginal vertex term and constructing the four-point scattering amplitude with two vertices. 
However, the situation is the same as the case with the four point vertex.
All three-point marginal operators can be expressed by 
\begin{eqnarray}
(\Box \phi)^2 \phi, \hspace{1cm} (\Box \phi)(\partial_\mu \phi)^2. \label{3pt}
\end{eqnarray} 
The nonrenormalizable term is the former, but $\Box$ operator operates different $\phi$. 
In each vertex of tree-level four-point amplitudes, only one line is internal, and thus we have still one $\Box$ operator in the external line.
Although it would be interesting to check the six-point amplitude, 
the calculation is complicated. 
Meanwhile, we can see the relation by the four-point vertex in the simple model that we introduce in Sec.\ref{HDSFT}. 
Therefore, we do not analyze the nonrenormalizable interaction terms appearing in (\ref{3pt}) and (\ref{4pt}).

\section{Quantization}\label{Quantization}

We show the quantization for fields $\psi_1$, $\psi_2$ and $\sigma$,
The quantization for canonical fields $\psi_1$ and $\sigma$ is done in the usual manner,
\begin{eqnarray}
&&\psi_1 = \int \frac{d^3p}{\sqrt{(2\pi)^3 E^{(1)}}} \left\{ a_1({\bf  p}) e^{ipx}+a_1^\dagger({\bf  p}) e^{-ipx} \right\}, \hspace{1cm} \left( E^{(1)}:=\sqrt{{\bf p}^2 +m_1^2}(=-p^{(1)}_0) \right), \\
&&\sigma = \int \frac{d^3p}{\sqrt{(2\pi)^3 E^{(\sigma)}}} \left\{ a_\sigma({\bf  p}) e^{ipx}+a_\sigma^\dagger({\bf p}) e^{-ipx} \right\}, \hspace{1cm} \left( E^{(\sigma)}:=\sqrt{{\bf p}^2 +m_\sigma^2}(=-p^{(\sigma)}_0) \right).
\end{eqnarray} 
The commutation relations of the creation and annihilation operators for the positive norm fields $\psi_1$ and $\sigma$ are
\begin{eqnarray}
\left[  a_1({\bf  p}),  a_1^\dagger({\bf  k}) \right]= \delta^3({\bf p}-{\bf k}) , 
\hspace{1cm} 
\left[  a_\sigma({\bf  p}),  a_\sigma^\dagger({\bf  k}) \right]= \delta^3({\bf p}-{\bf k}) .
\end{eqnarray}

The quantization of  $\psi_2$ is unsettled due to the negative sign of the kinetic term. 
We have two natural choices to define creation and annihilation operators for $\psi_2$,
\begin{eqnarray}
\psi_2 = \int \frac{d^3p}{\sqrt{(2\pi)^3 E^{(2)}}} \left\{ a_2({\bf  p}) e^{ipx}+a_2^\dagger({\bf  p}) e^{-ipx} \right\},\hspace{1cm} \left( E^{(2)}:=\sqrt{{\bf p}^2 +m_2^2}(=-p^{(2)}_0) \right), 
\end{eqnarray} 
or
\begin{eqnarray}
\psi_2 = \int \frac{d^3p}{\sqrt{(2\pi)^3 E^{(2)}}} \left\{ \tilde a_2^\dagger ({\bf  p}) e^{ipx}+ \tilde a_2({\bf  p}) e^{-ipx} \right\},\hspace{1cm} \left( E^{(2)}:=\sqrt{{\bf p}^2 +m_2^2}(=-p^{(2)}_0) \right).
\end{eqnarray} 
With the former definition, the creation operator creates a positive energy particle, but the sign of the commutation relation is flipped, $\left[  a_2({\bf  p}),  a_2^\dagger({\bf  k}) \right]= -\delta^3({\bf p}-{\bf k})$. 
On the other hand, the latter definition gives the usual commutation relation, $\left[ \tilde a_2({\bf  p}), \tilde a_2^\dagger({\bf  k}) \right]= \delta^3({\bf p}-{\bf k})$, but particle states have negative energy. 
We often choose the latter definition for physical fields (especially in cosmology), 
because the flipped sign of kinetic term would mean the negativity of kinetic energy and 
any state satisfies the positivity of the norm. 
However, the negative energy leads to the instability of the vacuum state (the ghost instability), and its decay rate is indeed infinity!
Thus, the perturbative approach breaks down.%
\footnote{
If Lorentz symmetry is broken, the decay rate of vacuum can be mild and we would be able to live with the ghost modes
\cite{Izumi:2008st,Izumi:2007pb}.
}
The renormalizability of the theory with higher order derivative is discussed with the former definition. 
Although it is unclear how to interpret the negative norm state, 
without the negative energy state the perturbative approach is applicable because of the absence of the ghost instability.
Therefore, we choose here the former definition.

The particle states are usually expressed by the  covariant creation operators, 
\begin{eqnarray}
A^\dagger_i ({\bf p}) = a^\dagger_i ({\bf p}) (2\pi)^{3/2} \sqrt{2 E^{(i)} ({\bf k})} \hspace{1cm} \left( i=1,2,\sigma \right),
\end{eqnarray} 
as
\begin{eqnarray}
|{\bf p}_{1_1},\cdots ,{\bf p}_{1_l};{\bf p}_{2_1},\cdots ,{\bf p}_{2_m};{\bf p}_{\sigma_1},\cdots ,{\bf p}_{\sigma_n}\rangle :=
A^\dagger_1 ({\bf p}_{1_1}) \cdots A^\dagger_1 ({\bf p}_{1_l})A^\dagger_2 ({\bf p}_{2_1})\cdots A^\dagger_2 ({\bf p}_{2_m})A^\dagger_\sigma ({\bf p}_{\sigma_1})\cdots A^\dagger_\sigma ({\bf p}_{\sigma_m}) |0\rangle.
\end{eqnarray} 
Based on the covariant creation operators $A^\dagger_i ({\bf p})$,
the usual Feynman rule gives the scattering amplitude. 
Then, the norm of one particle state becomes 
\begin{eqnarray}
\langle 0| A_i ({\bf k}) A_j^\dagger ({\bf p})|0 \rangle = \eta_i (2\pi)^3 2 E^{(i)} \delta^3({\bf k}-{\bf p}) \delta_{ij}, \label{innerP}
\end{eqnarray} 
with
\begin{eqnarray}
\eta_1=1=\eta_\sigma, \hspace{1cm} \eta_2=-1.
\end{eqnarray} 
The normalization factor $C_X$ for particle states defined in eq.(\ref{defCX}) is fixed by the above equation (\ref{innerP}).

\section{Derivation of (\ref{op3})}\label{Doop3}

Here, we show the derivation of (\ref{op3}) from the right hand side of inequality (\ref{op2}). 
Three of four delta functions represent the momentum conservation law, and there each component of ${\bf p}_j$ appears linearly. 
Therefore, the integration with respect to ${\bf p}_j$ can be easily done and in the center of mass frame 
it gives the replacement of ${\bf p}_j$ with $-{\bf p}_i$ in the integrand. 
The remaining delta function, namely the energy conservation, can be deformed as follows,
\begin{eqnarray}
\delta(E_1+E_2-E_i-E_j)& =&\delta\left(\sqrt{|{\bf p}_1|^2 + \mu_1^2}+\sqrt{|{\bf p}_1|^2 + \mu_2^2}-\sqrt{|{\bf p}_i|^2 + \mu_i^2}-\sqrt{|{\bf p}_i|^2 + \mu_j^2}\right) \nonumber\\
&=&
\frac{E_iE_j}{|{\bf p}_1|(E_i+E_j)} \delta \big(|{\bf p}_i|-|\hat{{\bf p}}_i| (|{\bf p}_1|)\big)\nonumber\\
&=&\frac{E_iE_j}{|{\bf p}_i|(E_1+E_2)} \delta \big(|{\bf p}_i|-|\hat{{\bf p}}_i|(|{\bf p}_1|)\big),\label{EC}
\end{eqnarray} 
where $\mu_k$ ($k=1,2,i,j$) shows the mass corresponding to the particle with momentum ${\bf p}_k$.
Taking the spherical coordinate for the momentum space of ${\bf p}_i$, $d^3{\bf p}_i$ is written as 
$|{\bf p}_i|^2 \sin \theta_p d|{\bf p}_i| d\theta_p d \phi_p$. 
Substituting this and eq.(\ref{EC}), integrating with respect to $\phi_p$, and ignoring unimportant numerical factor, 
we can obtain the form ($\ref{op3}$).

\section{Calculation of the scattering amplitudes } \label{calSA}

In Sec.\ref{SFMHOD} we have considered five amplitudes of two-two scattering of $\psi_1, \psi_2, \sigma$; (\ref{3a}) - (\ref{3e}). The momentum assignment of scattering is as 
\begin{eqnarray} 
{p}_1 + {p}_2  \to {p}_3 + {p}_4 \label{E1}
\end{eqnarray}
for all five processes. Here $p_i$ are four-momenta. The amplitudes are denoted by 
\begin{eqnarray}
{\cal M}\big(\psi_1({\bf p}_1)\psi_1({\bf p}_2)\to\psi_1({\bf p}_3)\psi_1({\bf p}_4)\big) \label{E2}
\end{eqnarray}
for the process (3a), and similarly for the other four. After taking account of the momentum conservation, we write 
\begin{eqnarray}
{\cal M}\big(\psi_1({\bf p}_1)\psi_1({\bf p}_2)\to\psi_1({\hat{\bf p}}_3)\psi_1({\hat{\bf p}}_4)\big) \label{E3}
\end{eqnarray}
similarly for other processes. Here $\hat{\bf p}_i$ is the momentum satisfying both of the on-shell condition and the energy-momentum conservation. The momentum variables ${\bf p}_1,.., {\bf p}_4$ are often omitted when it is understood. Through this paper we take the center of mass frame, hence 
\begin{eqnarray} 
{\bf p}_2 = - {\bf p}_1, \hspace{3mm}{\bf p}_4 = - {\bf p}_3  .\label{E4}
\end{eqnarray} 
The scattering angle is denoted by $\theta$.  $|\hat{{\bf p}}_i|$ is expressed in terms of $|{\bf p}_1|$, but its expression is different depending on the processes (\ref{3a}) - (\ref{3e}). The first three processes arise from the interaction term (\ref{4DNSI}) and hence their amplitudes are the same except for the all-over sign $\epsilon$, 
\begin{eqnarray}
{\cal M}=8 \epsilon \alpha\left[\left(p_{1}\cdot p_{2}\right)\left(p_{3}\cdot p_{4}\right)+\left(p_{1}\cdot p_{3}\right)\left(p_{2}\cdot p_{4}\right)+\left(p_{1}\cdot p_{4}\right)\left(p_{2}\cdot p_{3}\right)\right],\label{E5}
\end{eqnarray} 
where $\epsilon =1$ if the final state gives positive norm, otherwise $\epsilon = -1$. The last two processes arise from the interaction term (\ref{4thpsi-non}) and their amplitudes are given by 
\begin{eqnarray}
{\cal M}=4 \epsilon \alpha'\left(p_{1}\cdot p_{3}\right),\label{E6}
\end{eqnarray} 
where $\epsilon =1$ if final state gives positive norm, otherwise $\epsilon = -1$. However, the sign $\epsilon$ is immaterial in computing (\ref{op2}), (\ref{op3}), since the absolute values $|{\cal M}|^2$ are used there. 
  
\subsubsection{$\psi_1+\psi_1\to\psi_1+\psi_1$}
The two particles of the initial state and the two in the final state all have the same mass $m_1$ and hence $|\hat{{\bf p}}_i|=|{\bf p}_1|$. Using this in ${\cal{M}}$ of (\ref{E5}), we have 
  \begin{eqnarray}
{\cal M}\big(\psi_1({\bf p}_1)\psi_1(-{\bf p}_2)\to\psi_1(\hat {\bf p}_i)\psi_1(-\hat {\bf p}_i)\big) 
=8\alpha\big( (6+ 2\cos^2\theta)|{\bf p}_1|^4+8m_1^2|{\bf p}_1|^2+3m_1^4\big), \label{E7}
\end{eqnarray} 
The (\ref{op3}) of III.B now takes the form 
\begin{eqnarray} 
64 \alpha^2\frac{m_1^4|{\bf p}_1|^5}{E_1+E_2}\int_0^\pi \sin\theta (6+2\cos^2 \theta)^2  \left[ \left(\frac{|{\bf p}_1|}{m_1}\right)^4 + \frac{16}{6+2\cos^2\theta}\left( \frac{|{\bf p}_1|}{m_1}\right)^2 + {\cal O}\left( \left( \frac{m_1}{|{\bf p}_1|}\right)^0 \right) \right]     d \theta.\label{1111}
\end{eqnarray} 

\subsubsection{$\psi_1+\psi_1\to\psi_1+\psi_2$}
The expression of $|\hat{{\bf p}}_i|$ is slightly complicated containing $M_2 = {m_2}^2 - {m_1}^2$ 
\begin{eqnarray}
|\hat{{\bf p}}_i| &=& |{\bf p}_1| \sqrt{  1-\frac{M^2}{2|{\bf p}_1|^2}+\frac{M^4}{16|{\bf p}_1|^2(|{\bf p}_1|^2+m_1^2)}} \nonumber \\
&=& |{\bf p}_1| \left[ 1-  \frac{M^2}{4|{\bf p}_1|^2} + {\cal O} \left(\left(\frac{M}{|{\bf p}_1|}\right)^4\right) \right]    \,.\label{E9}
\end{eqnarray} 
Using this $|\hat{{\bf p}}_i|$ in ${\cal M}$ of (\ref{E5}), we have 
\begin{align}
&{\cal M}\big(\psi_1({\bf p}_1)\psi_1(-{\bf p}_2)\to\psi_1(\hat {\bf p}_i)\psi_2(-\hat {\bf p}_i)\big)\nonumber \\
&\quad
 =8\alpha\left( (6+ 2\cos^2\theta)|{\bf p}_1|^4+\left[8m_1^2-M^2(1+\cos^2\theta)  \right]|{\bf p}_1|^2+ m_1^2\left(3m_1^2-\frac{M^2}{ 2}\right)
-\frac{M^4}{8}+ \frac{M^4|{\bf p}_1|^2}{8(|{\bf p}_1|^2+m_1^2)} \cos^2\theta\right)\,. \label{E10}
\end{align} 
The (\ref{op3}) now takes the form,
\begin{eqnarray} 
&&-64 \alpha^2\frac{m_1^4|{\bf p}_1|^5}{E_1+E_2}\int_0^\pi \sin\theta (6+2\cos^2 \theta)^2  \nonumber \\
&&\qquad \times
\left[ \left(\frac{|{\bf p}_1|}{m_1}\right)^4 + \frac{1}{6+2\cos^2\theta}\left(16-\frac{M^2}{2m_1^2}(7+5\cos^2 \theta) \right)  \left( \frac{|{\bf p}_1|}{m_1}\right)^2 + {\cal O}\left( \left( \frac{m_1}{|{\bf p}_1|}\right)^0 \right) \right]     d \theta. \label{1112}
\end{eqnarray} 

\subsubsection{$\psi_1+\psi_1\to\psi_2+\psi_2$}
The two particles in the final-state have the same mass $m_2$ and hence $|\hat{{\bf p}}_i|$ is easily found, 
\begin{eqnarray}
|\hat{{\bf p}}_i| &=& |{\bf p}_1| \sqrt{ 1 - \frac{M^2}{|{\bf p}_1|^2}  } \nonumber \\
&=& |{\bf p}_1|\left[ \left( 1 - \frac{M^2}{2|{\bf p}_1|^2}\right) + {\cal O} \left(\left( \frac{M}{|{\bf p}_1|}\right)^4 \right)\right]\,, \label{E12}
\end{eqnarray} 
Using this $|\hat{{\bf p}}_1|$ in ${\cal M}$ of (\ref{E5}), we have
\begin{eqnarray}
&&{\cal M}\big(\psi_1({\bf p}_1)\psi_1(-{\bf p}_2)\to\psi_2(\hat {\bf p}_i)\psi_2(-\hat {\bf p}_i)\big)\nonumber \\
&&\qquad
 =8\alpha\left( (6+ 2\cos^2\theta)|{\bf p}_1|^4+\left[8m_1^2-2M^2(1+\cos^2\theta)  \right]|{\bf p}_1|^2+ m_1^2(3m_1^2-M^2)\right). \label{E13}
\end{eqnarray} 
The (\ref{op3}) now takes the form 
\begin{eqnarray} 
&&64 \alpha^2\frac{m_1^4|{\bf p}_1|^5}{E_1+E_2}\int_0^\pi \sin\theta (6+2\cos^2 \theta)^2  \nonumber \\
&&\qquad \times
\left[ \left(\frac{|{\bf p}_1|}{m_1}\right)^4 + \frac{1}{6+2\cos^2\theta}\left(16-\frac{M^2}{ m_1^2}(7+5\cos^2 \theta) \right)  \left( \frac{|{\bf p}_1|}{m_1}\right)^2 + {\cal O}\left( \left( \frac{m_1}{|{\bf p}_1|}\right)^0 \right) \right]     d \theta.\label{1122}
\end{eqnarray}      
We note that, if we set ${M}^2 =0$, all three amplitudes (\ref{E7}), (\ref{E10}) and (\ref{E13}) coinside. 

\subsubsection{$\sigma+\psi_1\to\sigma+\psi_1$}
Since the two final state particles are the same as the two initial ones, $|\hat{{\bf p}}_i| = |{\bf p}_1|$. Using this  $|\hat{{\bf p}}_i|$ in ${\cal M}$ of (\ref{E6}), we have 
 \begin{eqnarray}
{\cal M}\big(\sigma({\bf p}_1)\psi_1({\bf p}_2)\to\sigma(\hat {\bf p}_i)\psi_1(\hat {\bf p}_j)\big) =-4\alpha'\big( (1-\cos\theta)|{\bf p}_1|^2+m_\sigma^2\big).\label{E15}
\end{eqnarray} 
The  (\ref{op3}) now takes the form 
\begin{eqnarray} 
&&16 {\alpha'}^2\frac{|{\bf p}_1|^5}{E_1+E_2}\int_0^\pi \sin\theta  \left[ (1-\cos\theta)^2 + (1-\cos\theta) \frac{2m_\sigma^2}{|{\bf p}_1|^2}+ {\cal O}\left(\left(\frac{m_1}{|{\bf p}_1|}\right)^4\right) \right]     d \theta.\label{s1s1}
\end{eqnarray} 

\subsubsection{$\sigma+\psi_1\to\sigma+\psi_2$}
This process involves three unequal masses and $|\hat{{\bf p}}_i|$ in terms of $|{\bf p}_1|$ is more involved; we only show the high-energy approximation, 
\begin{eqnarray}
|\hat{{\bf p}}_1|= |{\bf p}_1|\left[ 1 - \frac{M^2}{4|{\bf p}_1|^2} + {\cal O} \left(\left(\frac{m_1}{|{\bf p}_1|}\right)^4\right)\right] . \label{E17}
\end{eqnarray} 
Using this $|\hat{{\bf p}}_i|$ in ${\cal M}$ of (\ref{E6}), we have 
\begin{eqnarray}
{\cal M}\big(\sigma({\bf p}_1)\psi_1({\bf p}_2)\to\sigma(\hat {\bf p}_i)\psi_2(\hat {\bf p}_j)\big) &=&4\alpha'\big( \sqrt{|{\bf p}_1|^2+m_\sigma^2}\sqrt{|\hat{{\bf p}}_i|^2+m_\sigma^2}-|{\bf p}_1||\hat{{\bf p}}_i| \cos \theta  \big). \nonumber\\
&=&
4\alpha' |{\bf p}_1|^2 \left[ (1-\cos\theta)\left( 1-\frac{M^2}{4|{\bf p}_1|^2}\right)  \frac{m_\sigma^2}{|{\bf p}_1|^2}+ {\cal O}\left(\left(\frac{m_1}{|{\bf p}_1|}\right)^4\right) \right].\label{E18}
\end{eqnarray}
The (29) now takes the form 
\begin{eqnarray} 
&&-16 {\alpha'}^2\frac{|{\bf p}_1|^5}{E_1+E_2}\int_0^\pi \sin\theta  \left[ (1-\cos\theta)^2 \left( 1-\frac{3M^2}{4|{\bf p}_1|^2}\right)  + \frac{2m_\sigma^2}{|{\bf p}_1|^2}+ {\cal O}\left(\left(\frac{m_1}{|{\bf p}_1|}\right)^4\right) \right]     d \theta.\label{s1s2}
\end{eqnarray}

\end{document}